# IT in Power Sector: A KPCL Implementation

**Pushpavathi T.P. and Shashi Kumar N.R. ,** *,CSE Dept.*
*DSCE, Visvesvaraya Technological University*

*Guide :* **Selvarani R** *, Associate Professor ,CSE Dept. DSCE*

## ABSTRACT

In this paper we investigate the extent of Information Technology penetration in Power sector, taking KPCL , Karnataka Power Corporation Ltd., a premier power generating , a state owned public sector organization as an example.

Any organization to flourish, adoption of Information Technology is inevitable in the days of fast changing technological advancements. It is not merely the investment on IT which helps but adoption of right IT solutions and the optimum use of the same does matter and becomes most critical. A strong infrastructure coupled with modern technical and management concepts has helped KPCL to meet the challenges of the rising energy demands of Karnataka.

A theoretical framework of the technology implemented in KPCL communication is discussed. This discussion is focused on the advantages of using the new technology (MPLS) over traditional systems which is in use and the related applications and services provided by the existing system. The issues among local connectivity and remote connectivity which are geographically distributed are addressed with the use of LANs and either VSAT or MPLS respectively. The contribution highlights the differences in varying levels of applications and services between VSAT and MPLS technology.

Finally a list of applications are presented that can benefit from the computerization in the enterprise segment.

## I INTRODUCTION

Karnataka, the eighth largest state in India. Rich in natural resources has the unique distinction of Asia's first hydro-electric power station established in 1902 at Shivasamudram on the banks of the river Cauvery. In fact Karnataka pioneering spirit in the field of power has been translated into several major milestones. Karnataka was the first to embark on Alternating Energy when Bangalore city's lighting scheme was completed. Karnataka was the first to have the longest transmission line in the world in 1902 from Shivasamudram to KGF covering a distance of 147 kms. Karnataka was also the first state in the country to conceive and set up a professionally managed corporation to plan, construct, operate and maintain power generation projects in the state, thus was the born of KPCL, Karnataka Power Corporation Ltd., [1]

It has been a long, rewarding journey. A journey fraught with trials and triumphs with challenges and achievements.

Today, KPCL takes great pride in the experience it has gathered, the expertise it has developed and the skills it has honed. Especially in the planning, investigation, design, execution and effective operation of large power projects.

### KPCL's Mission

The Mission of KPCL is to maximize power generation by identifying and developing opportunities in power generation while constantly upgrading technical competence and systems, developing human resources capabilities and their empowerment. Professionalism is the dominant ethos at KPCL. It strives to promote a work culture that brings out the best in its people. It is proud of its ability to attract and retain professionals of the highest caliber.

## Installed Capacity

KPCL today has an installed capacity of 4995 MW of hydel, thermal and wind energy, with 4000 MW in the pipeline. The 1470 MW Raichur Thermal Power Station located in Raichur dist is accredited with ISO 14001-2004 certification for its environment protection measures. From an industry vantage point, KPCL has raised the bar on the quality of deliverables and is constantly working at lowering the cost per megawatt - a commendable cost-value equation that has become a benchmark on the national grid. KPCL's stock in trade is industry proven - well-established infrastructure & modern, progressive management concepts and a commitment to excel, helping it to meet the challenges of the rising energy demands of Karnataka. The table 1 below shows the KPCL power projects in Karnataka. [1]

The leverage point of KPCL initiatives are its resource management strengths – right across planning, financing and project engineering. KPCL also has a high rating in terms of project completion and commissioning within the implementation calendar.

**KPCL's Financial Management :**

A turnover of Rs 16300 Million, operates at a profit of Rs 6920 Million and net profit of Rs 1420 Million for the year 1999-2000. These figures add up to, are strong financials. The figures for 2006-07 are: Sale of Energy Rs 3433.82 Crores, a total income of Rs 3750.38 Crores and total profit after taxes is Rs 322.32 Crores.[2]

**KPCL Power Projects in Karnataka**

| | Projects | Capacity (in MWs) |
|---|---|---|
| **A** | **Hydel** | |
| 1 | Cauvery River Basin Projects | 59.20 |
| 2 | Sharavathy Hydro Electric Project | 1469.20 |
| 3 | Kalinadi Hydro Electric Projects | 1225.00 |
| 4 | Varahi Hydro Electric Projects | 239.00 |
| 5 | Krishna River Basin Projects | 290.00 |
| 6 | Mini Hydro Electric Projects | 109.95 |
| | **Total Hydro** | **3392.35** |
| **B** | **Thermal** | |
| 1 | Raichur Thermal Power Station | 1470.00 |
| **C** | **Diesel Generating Station** | |
| 1 | Yelahanka DG Plant | 127.92 |
| **D** | **Wind** | |
| 1 | Kappatagudda Wind Farm | 4.55 |
| | **Grand Total** | **4994.82** |

**Consultancy**

KPCL today has the capability to undertake large scale power projects from concept to commissioning. It can also operate the plant on an EPC basis, with a host of exclusive auxiliary services. KPCLs Consultancy and Engineering Services Division, an offshoot of its core competency, offers its clients a wide spectrum of consultancy inputs across the complete cycle of power project development. It has the expertise in analysis and design of structures using STADPRO – 2006, NISA – Finite Element package, AUTOCAD – 2006, microstation and in house developed software packages for reservoir operation, Stability of Dams etc.,

## II IT INFRASTRUCTURE

KPCL is a highly IT Savvy company among the public sector units in the state at large and power sector in particular and has the distinction of being the first state owned company in the power sector to have established SHAKTINET- a satellite based communication network through VSATs , Very Small Aperture Terminals among all its power stations and Bangalore. Local Area Networks have been established at all Project locations and about 800 officials have been wired to corporate office with reliable VSAT back bone. All the activities of the corporation are computerized and KPCL has the unique distinction among power utilities of establishing voice and data communication between its projects and Bangalore through the new generation DAMA V-SATS on KPTCL's "VIDYUTNET".

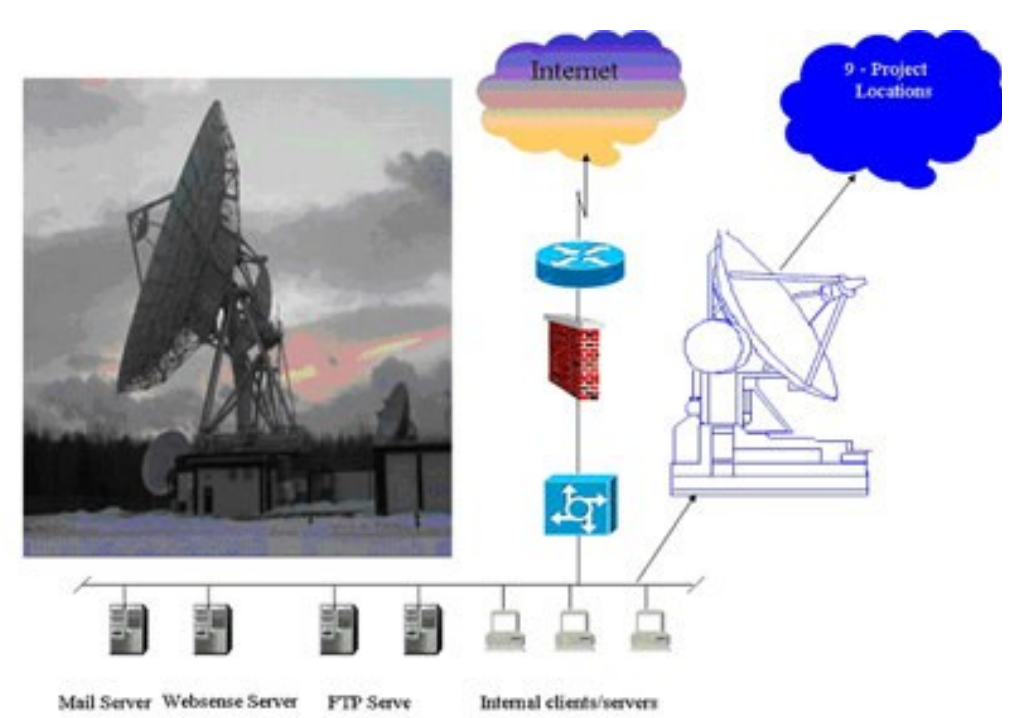

Figure 1 ShaktiNet

**ShaktiNet**

ShaktiNet is as sown in Figure 1 is a closed user group (CUG) satellite based telecommunication network of KPCL which employs state-of-the-art hybrid technology. The network connects the HUB station installed at LDC , Load Dispatch Centre of KPTCL with all the VSATs installed at all major power stations in the state of Karnataka.The uniqueness of the network is the deployment of VSAT technology to establish communication at remote project places where the availability of communication infrastructure is very critical and it is difficult to depend on the regular telephones.

**Configuration**: ShaktiNet has been implemented by using VSAT technology (Very Small Aperture Terminal) which is highly reliable and cost effective. The network supports both star and mesh configuration in Demand Assigned connectivity (DAMA technology).

The ShaktiNet consists of three major components viz,central HUB station at LDC, the satellite Insat-2C and ten remote terminals. The network operates in Extended C band using carrier frequencies of 6.725-7.025 GHz for unlinking and 4.5-4.8 GHz for down linking and uses transponder space segment of Insat-2C.

A bandwidth of 256 Kbps for data and 32 Kbps for voice totaling to 288 Kbps is being provided through Vidyuthnet. The satellite at the Central location at HO has 8 floating channels each of 32 Kbps for data and 4 voice channels each of 8 Kbps. These 8 floating data channels are being utilized by 10 project locations for applications such as mailing, Internet browsing, Financial Accounting, Integrated Inventory Management System etc

**Enterprise wide Area Network:**

KPCL has established a high speed Ethernet based Local Area Network at both its corporate office and at all project locations. Wireless LAN is also implemented at corporate office. KPCL Network via Satellite and MPLS is as shown in the Figure 2.The following services are available on Enterprise wide Network:

- Internet
- Email
- File Transfer
- Remote Login
- Project Management
- Intranet Connectivity

**Optical Fiber Connectivity**

Optical Fiber connectivity has been established at project locations to connect Power Houses and Office Complex at the following project locations:

- Jog office complex – Sharavathy Generating Station
- Jog Office complex – MGHE –Kargal
- Gerusoppa office complex – Gerusoppa Dam Power House
- Raichur Thermal Power Station – Office Complex

**Leased Line Connectivity**

KPCL has also established leased line connectivity to connect its different branch offices at Bangalore.

**New Initiatives**

Due to increased Data activities over the WAN, the bandwidth of 288 Kbps provided by Vidyuthnet (VSAT) was observed to be absolutely inadequate and KPTCL was not in a position to provide the higher bandwidth requirement of KPCL for its increased data activities over the WAN.

In order to meet the growing demands for bandwidth, it was decided to go with BSNL for the MPLS circuits which comprises of Fiber Optic Cable ring laid by them across all the District Headquarters in Karnataka and a smaller length of Copper Conductors from the Dist. HQ to the nearest telephone exchange at the desired project locations.

**MPLS**

Multi Protocol Label Switching adds to the capabilities of IP networks in several ways. Despite new capabilities, MPLS technology has much in common with ordinary IP networks.

Multi Protocol Label Switching (MPLS) is an extension to the existing Internet Protocol (IP) architecture, MPLS enables new support of new features and applications which include traffic engineering, IP virtual private networks, integration of IP routing and layer 2 or optical switching, and other applications.

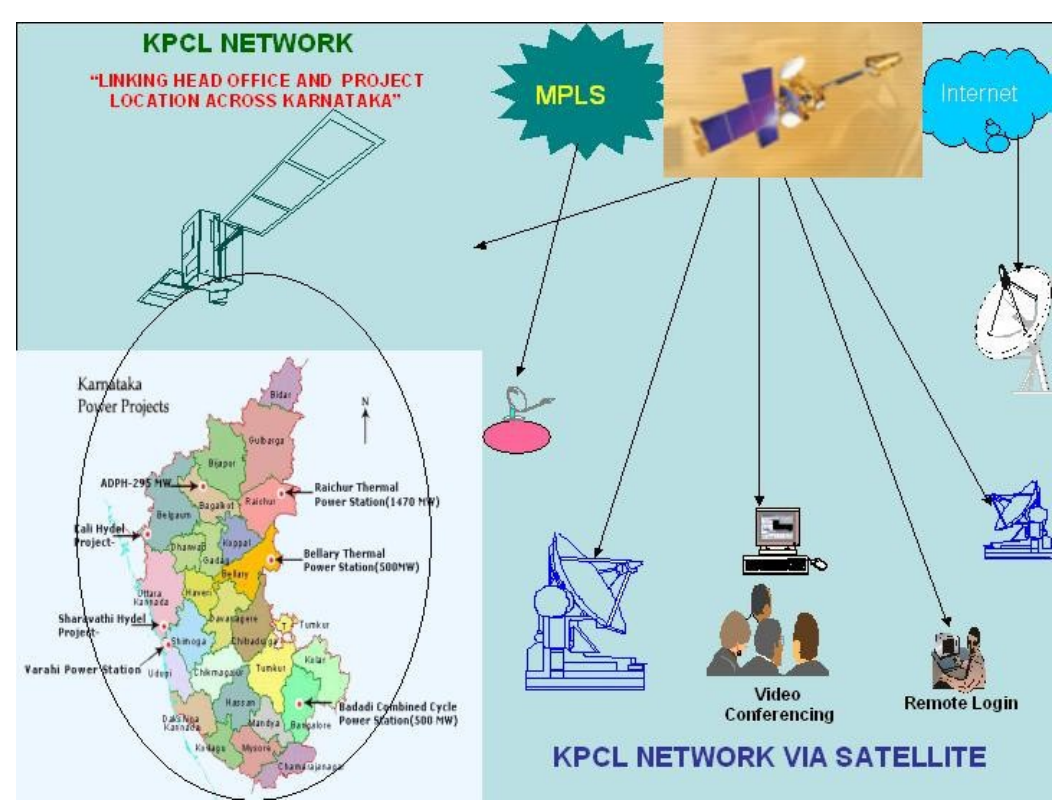

Figure 2 KPCL Network via Satellite and MPLS

MPLS networks use a superset of the protocols used in ordinary IP routers, and most MPLS networks are fully interoperable with ordinary IP equipment. The **P** protocol architecture, or IP stack, consists of a large number of protocols, a subset of which is used on IP routers, as opposed to hosts or servers. MPLS significantly extends the capabilities of IP networks by introducing support of virtual private networks and traffic engineering, and with integration of IP routing with various layer 2 switching technologies. MPLS achieves these new capabilities by extending the IP protocol architecture, rather than replacing it.

**MPLS Technology**

MPLS is a versatile solution to address the problems faced by present-day networks – speed, scalability, quality-of-service (QoS) management, and traffic engineering. It combines the speed and performance of packet-switched networks with the intelligence of circuit-switched networks to provide a best-of-breed solutions for integrating voice, video and data. Like circuit-switched networks, MPLS establishes the end-to-end connection path.

**MPLS Benefits**

- Cost reduction through data network convergence
- Integration of voice, video and data services
- New high-margin revenue opportunities through MPLS-based service offerings

**MPLS Applications**

- Improves packet-forwarding performance
- Supports network scalability
- Integrates IP and ATM in the network
- Builds interoperable networks

KPCL has established 2 Mbps MPLS network at the following projects for speedy voice and data communication:

- Bellary Thermal Power Station
- Raichur Thermal Power Station
- Nagjhari Power House
- Sharavathy Generating Station
- Varahi Underground Power House
- Bangalore Corporate House

KPCL MPLS provides following services:

- Video Conferencing
- Remote Administration
- Voice over IP

**Blade servers**

Blade servers are self-contained computer servers, designed for high density. Whereas a standard rack-mount server can exist with (at least) a power cord and network cable, blade servers have many components removed for space, power and other considerations while still having all the functional components to be considered a computer. A blade enclosure provides services such as power, cooling, networking, various interconnects and management—though different blade providers have differing principles around what should and should not be included in the blade itself. Together these form the blade system.[3]
KPCL has installed these blade servers from Hewlett Packard for the following applications:

- Integrated Inventory Management System
- Fuel Management System
- Email
- Finance applications
- HR application
- Establishment
- Gateways and Firewalls
- Web Security
- Trend Micro Centralized Anti Virus System

**Internet Connectivity**

VSNL is the Internet Service Provider (ISP) for KPCL for its internet connectivity. KPCL has got 2 Mbps connectivity from VSNL. Proxy server has been set up to cater internet connectivity for over 100 users.

## III SOFTWARE DEVELOPMENT

The computer services dept., of KPCL with fully dedicated engineers and software professionals has developed and implement all the software required for its business activities and is geared to offer consultancy services in the field of software development and working. KPCL has standardized Oracle database as a back-end and Developer-2000 which consists of Forms and Reports as a front-end tools. All applications are developed using these software tools. However some of the web based applications are developed using ASP (Active Server Pages), HTML (Hyper Text Markup Language) Some of the software developed in-house are:

- Itegrated Inventory Management System
- Human Resources Information System
- Fuel Management System
- Financial Accounting software
- E-Recruitment
- Establishment software
- Generation Management
- Medicine Procurement System

**Remote Data Acquisition System using GSM Technology**

The system is a GSM based data transmitter and receiver system which is capable of receiving and displaying water level data from maximum of 10 WLD (Water Level Data) transmitters through SMS over GSM network. Each transmitter's data gets time slice of 10 seconds to display data in a continuous manner. The system is over-the-air configurable with a command SMS from a GSM mobile from any where in the world.

Commands are provided to configure and communicate with the system over GSM network through SMS as well as thru computer hyper terminal. Commands provided to set WLD transmitter IDs, to set Higher and Lower water limits for an ID, to set system for factory setting, to request for configuration parameters of the system.

Received water level data which contains ID of the WLD transmitter, Time, Date and water level either in feet or in meter is displayed on the respective displays provided. Visual and Audible alerts are provided to intimate the user if received water level goes beyond higher and lower water limits set for an ID. The system also has serial interface with the host computer for logging received water level data in computer.

**Project Management**

KPCL has been using Prima Vera software product for state-of-the-art project implementation and monitoring techniques. It also extensively uses CAD and CAM in design working.

## IV SERVICES

**Video Conferencing**

KPCL has established Video Conferencing facility at following projects:

- Bellary Thermal Power Station
- Raichur Thermal Power Station
- Nagjhari Power House
- Sharavathy Generating Station
- Varahi Underground Power House
- Bangalore Corporate House

Video Conferencing services will connect the Corporate office at Bangalore to all other remote locations listed above. The VC equipments at Corporate office will have Multipoint Conferencing unit which facilitates group video conferencing from a point to possible many multipoint. The cameras used have Pan, tilt and zoom facilities connected to digital projector. These VC equipments can be connected to Video camera, PC and Visualizers so that video , powerpoint presentations and slides can be displayed respectively to other end.

**Voice over Internet Protocol**

VoIP technology leverages the connectionless features of TCP/IP in which packets can take different paths between end-points and all paths are shared by packets from different transmissions. Special headers are used to give voice packets priority over data.KPCL has implemented VoIP as it is cheaper than PSTN and it can save money on long distance calls

**KPCL Website**

KPCL has developed and hosted its own website and is available at the address www.karnatakapower.com.

This website gives details about the following:

- Mission
- Power Projects
- On-going and Future Projects
- Tender Notifications
- IT Infrastructure
- Awards and Achievements
- Environment Management
- Consultancy

**E-mail Connectivity**

KPCL has installed and configured Microsoft Exchange Server for its Email needs. All its ten project locations and corporate office have E-mail facility.

## V NETWORK SECURITY

Today's employee computing environment offers access to powerful applications and downloads, which can introduce new security challenges for corporate IT. With the nature of the employee workplace becoming increasingly more interconnected and mobile, the frequency and cost of the resulting security intrusions can be high.
Today's organizations face an increasing number of threats across their network – at the perimeter, across the LAN and WAN, and at endpoints. Protecting them all can result in security sprawl, an ever-increasing set of product that is hard to manage and result in inconsistent security.

**Websense**

Spyware applications 'spy' on user activities and relay collected information back to the originating computer. These programs are far from benign, as they are designed to scan the systems,monitor activity and relay confidential information to other computers. Once considered only an annoyance,spyware has evolved from a nuisance to a malicious threat. Preventing spyware from infiltrating an organization requires security measures at multiple points on the network, gateway and endpoint. Websense software is used in KPCL to combat this growing problem.

The Websense Solutions: Websense Web Security Suite is a leading internet security solution installed in KPCL that protects organization from spyware, malicious mobile code and phishing and pharming attacks. It also blocks spyware and key-logger backchannel communications from ever reaching their host servers. Websense Web Security Suite blocks sites that host spyware applications and are used for drive-by installations; malicious sites that exploit browser vulnerabilities or are used to post information to or from spyware and sites that host CGI mail parsers from which information captured by spyware could be extracted. Web Security Suite scans and classifies over 450 Million websites per week for malicious activity. Customer databases are then automatically updated with the new classifications. The suite identifies new threats quickly, resolving the issue and decreasing threat exposure time. Real-time security updates are available within minutes of the discovery of a new high-risk threat with no administrative intervention is required.[4]

**Firewalls**

Every PC accessing your enterprise network is a target for rapidly proliferating worms, penetration attacks, Trojan horses, spyware and other exploits. Reactive, signature-dependent technologies such as antivirus and intrusion detection can no longer be trusted to stop the newest variations of these threats. There is also a growing need within large enterprises to extend networks, applications and corporate databases. This need has resulted in large, complex networks that are hard to manage and secure. Organizations are required to deploy firewalls, VPN and intrusion prevention devices in front of each network segment to achieve comprehensive security.

**Checkpoint's VPN-1 Internet Gateway:**

VPN-1 Power VSX is a high speed multipolicy virtualized security solution designed for large-scale enterprise environments like data centers and campus networks. KPCL has installed VPN-1 from Check Point Software Technologies. VPN_1 Power VSX provides comprehensive protection to multiple networks or VLANs within complex infrastructures securely connects them to shared resources like the internet and allows each of them to interact with each other safely, while providing centralized management.[5]

VPN-1 VSX is a virtualized security gateway that allows virtualized enterprises to create up to 250 virtual systems – firewall, VPN and intrusion prevention functionality within a virtual environment. It features:

- Virtualized security including firewall , VPN and Intrusion prevention
- Virtualized Network Environment

**Anti-Virus Protection System**

KPCL has placed worldwide renowned and reputed Trend Micro Office Scan suite , Spam protection service and Damage Cleanup services to centrally manage and control virus , worms threats.

## VI CONCLUSION AND FUTURE WORK

An overview of the IT in the KPCL organization has been provided. Issues related to Network like VSAT, MPLS, Leased line have been outlined. Results for this IT framework, applications and services have been identified.

Experience of the above computerization shows that availability of the bandwidth in VSAT technology is limited compared to MPLS technology. However the guaranteed uptime of the VSAT network is considerably high compared to MPLS as the last mile is still on the copper cable that tends to breakdown frequently. Otherwise the applications and services that can be catered on MPLS are of significant importance over VSAT because of the high bandwidth availability in MPLS.

Hence future work with KPCL is to find a solution to the last mile problem in MPLS to address the issue relating to breakdown of the network and to make it high uptime.

The right IT solutions can help to lower Total Cost of Ownership (TCO) and increase Return On Investment (RoI). KPCL has set for itself very high standards and firmly believes that "It is not the IT Infrastructure of a few crores that needs to be managed but the business that gets transacted over the network that needs to be managed "

## BIOGRAPHIES

Pushpavathi T.P. [ace/pushpaOyahoo.co.in) is a final year student of Master of Technology in Computer Network Engineering in Dayanand Sagar College of Engineering, Bangalore affiliated to Visvesvaraya Technological University. She completed her Bachelors Degree in Computer Science and Engineering from University of Mysore.

Shashi Kumar N.R. [chandravalliOyahoo.com, shashikumarOkarnatakapower.com) is a final year student of Master of Technology in Computer Network Engineering in Dayanand Sagar College of Engineering , Bangalore affiliated to Visvesvaraya Technological University. He completed his Bachelors Degree in Computer Science and Engineering from University of Mysore.